# Energy Autonomous Wearable Sensors for Smart Healthcare: A Review

Abhishek Singh Dahiya[1*], Jerome Thireau[3], Jamila Boudaden[4], Swatchith Lal[5], Umair Gulzar[6], Yan Zhang[6], Thierry Gil[2], Nadine Azemard[2], Peter Ramm[4], Tim Kiessling[4], Cian O'Murchu[5], Fredrik Sebelius[7], Jonas Tilly[7], Colm Glynn[8], Shane Geary[8], Colm O'Dwyer[5,6,9,10], Kafil M. Razeeb[5], Alain Lacampagne[3], Benoit Charlot[1*], Aida Todri-Sanial[2*]

[1]*IES, Université de Montpellier, CNRS, Montpellier, France*
[2]*LIRMM, Université de Montpellier, CNRS, Montpellier, France*
[3]*PhyMedExp, CNRS 9214, INSERM U1046, Université de Montpellier, France*
[4]*Fraunhofer Research Institution for Microsystems and Solid State Technologies EMFT, Silicon Technologies and Devices, Hansastrasse 27d, 80686 Munich, Germany*
[5]*Micro-Nano Systems Centre, Tyndall National Institute, Lee Maltings, Dyke Parade, Cork, T12 R5CP, Ireland*
[6]*School of Chemistry, University College Cork, Cork, T12 YN60, Ireland*
[7]*Novosense AB, Scheelev. 17, 223 70 LUND, Sweden*
[8]*Analog Devices International, Raheen, Limerick*
[9]*AMBER@CRANN, Trinity College Dublin, Dublin 2, Ireland*
[10]*Environmental Research Institute, University College Cork, Lee Road, Cork T23 XE10, Ireland*

***Corresponding author(s) email: abhisheksingh.dahiya@ies.univ-montp2.fr; benoit.charlot@um2.fr; aida.todri@lirmm.fr**

**Abstract**

Energy Autonomous Wearable Sensors (EAWS) have attracted a large interest due to their potential to provide reliable measurements and continuous bioelectric signals, which help to reduce health risk factors early on, ongoing assessment for disease prevention, and maintaining optimum, lifelong health quality. This review paper presents recent developments and state-of-the-art research related to three critical elements that enable an EAWS. The first element is wearable sensors, which monitor human body physiological signals and activities. Emphasis is given on explaining different types of transduction mechanisms presented, and emerging materials and fabrication techniques. The second element is the flexible and wearable energy storage device to drive low-power electronics and the software needed for automatic detection of unstable physiological parameters. The third is the flexible and stretchable energy harvesting module to recharge batteries for continuous operation of wearable sensors. We conclude by discussing some of the technical challenges in realizing energy-autonomous wearable sensing technologies and possible solutions for overcoming them.

## 1. Introduction

The Internet-of-Things (IoT) is a network of interconnected devices that are playing an important role to improve the quality of individuals' life by providing a "smart" environment. IoTs have found commercial success in application areas such as smart city, transport and home. Healthcare is one of the crucial sectors where IoTs can provide individuals' safety and comfort by continuous health monitoring. In this regard, research efforts are dedicated to the "Internet-of-Medical-Things (IoMT)", linking billions of wearable devices/sensors into a communication network allowing patient-doctor communications periodically or real-time[1]. **Figure 1** describes the concept of IoMT where wearable connected devices collect health data and transfer to healthcare service providers[1–6]. IoMTs provide the opportunity for medical practitioners to monitor risk patients. On-going research efforts are focused on development of multi modal wearables that can be easily integrated in clothing[7–9], wrist watches and bands[10–13], contact lenses[14–18], tattoo-like sensory skin patches (electronic-skin)[19–25] etc. to enable around-the-clock vital signs monitoring. The battery is a significant part of an IoMT to supply continuous power to the sensors, however it is also bulky, with a limited lifespan and requiring periodic replacement[26]. For long-time wearable IoMT, energy autonomy is critical. Depending on the number of sensors, size and complexity of electronic circuitry, power requirement for wearables could range from 1 to 100 $\mu$W. Such demand could be met with smart nanogenerators (NGs), and that is why several green energy sources are being investigated[27], as the possibility to develop "symbiotic" devices[28].

An Energy Autonomous Wearable Sensor (EAWS), with low power consumption and a flexible package is indispensable for the successful deployment of IoMTs. In literature, several surveys can be found on EAWS for personal healthcare. For instance, extensive surveys presented in ref[2,3,5,29,30] cover aspects of wearable sensors, whereas in ref[1] focus is given to IoT for healthcare monitoring covering aspects of short- and long-range communications standards and cloud technologies. This paper provides a thorough review of wearable autonomous sensing system with a focus on sensors, batteries and energy harvesting. In Section 2, we describe the various types of sensors and their sensing mechanisms suitable for health monitoring. In Section 3, we describe the recent developments in flexible energy storage for wearables. In Section 4, we present the recent developments in energy harvesting for enabling long-lasting energy autonomy. We conclude the review in Section 5, where we provide an overview of the main challenges and viable solutions for future IoMTs.

## 2. Wearable sensors

Here, we present various sensors and their sensing principle for wearable devices. Different types of wearable sensors are reported to-date such as mechanical, biopotential electrode, optical and biochemical sensors to monitor different physiological vital signals. In the following subsections, we present a survey of mechanical and biochemical sensors.

### 2.1 Wearable mechanical sensors

Mechanical sensors such as strain and pressure sensors are developed and mounted on skin or integrated into clothing. Such sensors work on transduction mechanisms such as piezoresistive, capacitive, piezoelectric, and triboelectric. A representative summary of the performances of mechanical strain/pressure wearable sensors reported is provided in Table 1 (piezoresistive and capacitive) and Table 2 (piezoelectric and triboelectric). Each of them is discussed in details in the following section.

#### 2.1.1 Piezoresistive sensors

Piezoresistive sensors are based on the change of electrical properties of the material when subjected to mechanical deformation. This electromechanical response of the material is called piezoresistive effect or resistive change. Piezoresistive sensor is often named as a resistive strain sensor or simply strain gauge. The piezoresistive effect has been widely exploited using many different deformable materials for the detection of physiological body signals due to many advantages such as simple device design, high sensitivity and simple readout circuits[20,31–

[41]. The important requirements for resistive strain sensors to be employed for wearable applications are: (i) high flexibility and stretchability to absorb more strain with body movements, (ii) compactness and small size, (iii) high sensitivity, and (iv) biocompatibility. The strain sensitivity is typically characterized with a gauge factor (GF) as $GF = \frac{\Delta R}{R_0} / \frac{\Delta L}{L_0}$ where $\Delta R$ is the change in resistance, $R_0$ is the unstrained resistance, $\Delta L$ is the change in device length and $L_0$ is the initial length of the device. Simplest of the resistive strain sensors are metal strain gauges. They are typically a meander shaped metal wire patterned on a flexible polyimide film and sensitive to bending[42,43]. To transfer the technology over stretchable substrate (to be compatible with wearable application), a thin-film or patterned shaped metallic conductor is deposited over a silicone elastomer, for example polydimethylphenylsiloxane (PDMS). When the device is stretched, the geometrical change induces a change in dimensions, thus a change in electrical resistance. The sensitivity or the GF of such devices is low, typical values are between 1-5.[43] To improve the sensitivity of metal based stretchable strain sensors, microcracking of the metallic thin-films has been shown promising results.[32,33] For example, Kang et al. reported nanoscale crack junctions in Platinum (Pt) metal thin films, inspired by the crack-shaped slit sensory organs of spiders, with very high sensitivity.[33] A GF of 2000 over a range of 0–2% is achieved, allowing for detection of physiological signals such as speech patterns and heart rate. However, the durability and stretchability of such a device is limited, showing signal degradation at about 500 cycles of 2% strain. The microcracking-based approach may result in achieving a high GF, but device stability is a big issue to be resolved.

To address this issue, a new class of materials called ‘nanocomposite materials’ are being investigated.[44] This new class of material is a mixture of viscoelastic matrix (polymer) and conductive fillers (i.e., two-dimensional (2D) or 3D scaffolds of 1D nanomaterials) which are the most feasible material choice to meet the requirements of wearable mechanical sensors. These materials provide accurate and reliable sensing while enabling movement and comfort for users. This class of material is unique as it retains many desirable features of polymers (flexibility, biocompatibility, processability) yet, it adds electrical conductivity and piezoresistive property of the 1D nanomaterials to design lexible sensors with skin-line conformability and stretchability.[45] Various polymers such as poly (methyl methacrylate) (PMMA) polycarbonate (PC), poly (ethylene) (PE), poly (L-lactide) (PLLA), etc. have been incorporated with nanofillers to construct strain sensors, which are capable of holding larger tensile strain than conventional metallic strain gauges.[46] Compared with above polymers, silicone-based polymers such as PDMS, dragon-skin© (Smooth-On, Inc.) and Ecoflex© (Smooth-On, Inc.) own superior mechanical elasticity as it can easily endure over 100% of tensile strain without any structural failure[47,48], making them an ideal choice for large-range strain sensing applications.

Carbon nanotubes (CNTs), as a filler material, offer superior sensitivity owing to their remarkable electrical and mechanical properties.[6,34,49–51] The tunability of the physical properties of these nanomaterials is one of the most promising features for their implementation as a versatile sensing platform. Numerous experimental, numerical, and theoretical studies have been made on the electrical properties of these CNT/polymer nanocomposites. The results suggest that the electrical conductivity is generally attributed to the formation of conductive CNT percolation networks within the CNT/polymer nanocomposite. The networks depend on the electronic band structure, the tunneling resistance at crossed CNT junctions, and the morphology of the CNT percolation networks[52–55]. The principle of strain sensing for such devices is schematically shown in Figure 2a. In general, a low volume fraction of conductive nanomaterials results in high sensitivity but relatively low stretchability, whereas a high fraction of conductive nanomaterials results in higher stretchability but lower sensitivity.[41] Studies have also shown that the processing parameters, such as mixing speed, mixing time, resin temperatures and the viscosity at mixing, CNT aspect ratio (length/diameter), and curing temperatures, could influence the formation of the conducting network significantly, leading to a significant variation of electrical properties of CNT nanocomposites. One of the most important parameters is also the filling fraction of CNT in the polymer. Various studies have shown that the resistivity can be reduced to ~150Ω·cm (or $10^{-1}$ S/m) at a CNT content of 1.0 wt %. This value is still not sufficient for use in electronic devices as at least electrical conductivity of 1 S/m (or~100Ω·cm) is required to obtain meaningful ECG shape curves for precise detection of P, QRS and T waves.[56] Alternative to CNTs, metallic nanomaterials have good and electrical properties and high sensitivity (GF is easily achieved twice or even more times than CNTs)[57,58].

Meanwhile, different methods have been experimented to develop highly sensitive resistive sensors, such as pre-stretching, 3D printing, and microfluidic techniques.[41,59–63] For instance, Muth et al. used 3D printing to fabricate sensitive strain sensors with conductive carbon grease materials to obtain a high stretchability of 400%.[59] It is difficult to construct a high sensitivity strain sensor with improved stretchability. Ideally, a sensing device should have a large stretchability as well as a high GF to measure the change in the electrical resistance upon deformation. More recently, an innovative yet simple device fabrication approach is reported to show an enhanced strain sensing performance. In this fabrication approach, networks of 1D materials, such as CNTs or silver (Ag) nanowires (NWs), on top of a stretchable substrate are realized in a micro/macro channels ,which offers an alternative way to realize stretchable strain sensors with high GF.[41,49] Using this fabrication strategy, Han et al.,[41] created a NW-based strain sensor by hybridizing brittle metal nanowires (Ag NWs, Copper NWs, or CNTs) and a conductive organic solution, and poly(3,4-ethylenedioxythiophene):polystyrene sulfonate (PEDOT:PSS). The nanowire microfluidic hybrid (NMH) strain sensors exhibit record-high values for both the GF and stretchability. This is because the advantages of both materials have been coupled. The brittleness of the NWs endows the strain sensor with an outstanding sensitivity under a small stretchable load, and the conductive organic solution significantly enhances the deformation endurance of the device.

The nanocomposite material based piezoresistive sensors exhibit large stretchability and high sensitivity, but often suffer from drawbacks such as nonlinear response, large hysteresis, and irreversibility[34,64]. These disadvantages can be explained as nanomaterials cannot completely come back to their initial position when unloaded of the applied strain.

### 2.1.2 Capacitive sensors

As an alternative approach to piezoresistive sensing technology, capacitive-based sensors offer some advantages, such as higher linearity, less hysteresis, and fast response time, which are important parameters when sensors are intended to be used in real-life scenarios.[65–71] The simplest example of mainstream capacitive sensors is the parallel-plate configuration, as it is easy to construct and straightforward to model (Figure 2b). The capacitive change is governed by the classic equation as $C = \varepsilon A/d$ where ε is the permittivity of the cavity between two plates, and A and d represent the overlap area and distance between two plates, respectively. The change in device capacitance with the applied external force/pressure, can be measured either using a passive capacitor[65–68] or through modifying the response curve of an active component, such as using field-effect transistors (FET)[70].

Inorganic nanomaterials and conducting polymers, including CNTs, dinaphtho[2,3-b:29,39-f] thieno[3,2-b] thiophene (DNTT) and rubrene, have been employed for constructing flexible FETs-based tactile sensors.[36,70,72] For example, Schwartz et al. have demonstrated a new type of an active pressure sensor using the microstructured PDMS as a gate dielectric for rubrene single crystal FET. The applied pressure induces a change in the capacitance of the compressed microstructured PDMS dielectric layer, which transduces as the drain-source current modulates[70]. The fabrication of FET-based capacitive sensors, is however challenging as it is usually based on a complex and expensive process, which is difficult to scale-up for commercialization[72]. For passive capacitive strain sensors, silicone-based polymers and parallel plate capacitor arrangement are mostly exploited to build the sensing structures. In these devices, a soft dielectric silicone-based elastomer is sandwiched between conductive flexible plates. New electrode nanomaterials such as Ag NWs[73–75] and CNTs[76] have been exploited to improve the flexibility and stretchability of capacitive strain sensors.

Considerable research efforts, inspired by nature, have also been made to develop sensitive and stretchable capacitive pressure sensors.[65,71] Example of this approach can be seen in the work reported by Kang et al.[71] where they have demonstrated a facile approach to fabricate highly sensitive capacitive pressure sensors based on a sponge-like structure of PDMS thin-film dielectric layer, inspired by the pore structures of *spongia officinalis*. The morphology of the porous structured dielectric layer in the fabricated pressure sensors is controlled by changing the pore sizes. These flexible pressure sensors are capable of not only detecting extremely low pressures, but also enabling real-time tactile sensing applications. Such porous structured pressure sensor with 6 µm pore diameter exhibited high sensitivity (0.63 $kPa^{-1}$), high stability over more than 10 000 cycles of applied pressure, and fast response time (in the milliseconds range).

Piezoresistive effect is mainly exploited to measure strain, whereas capacitive effect is used to measure pressure. Both piezoresistive and capacitive based strain/pressure sensors are gaining lot of interest as they provide high sensitivity with simple device design and readout circuits. However, power consumption is a critical issue in the implementation of these sensors for EAWS. Research efforts are on-going to realise ultra-low power sensors to extend the life of rechargeable batteries. Meanwhile, mechanical sensors based on piezoelectric and triboelectric transduction phenomenon are gaining interest due to the possibility to realize self-powered sensors with negligible power consumption over flexible and stretchable substrates. The following sections will discuss on some of the best reported sensors in literature.

### 2.1.3 Piezoelectric sensors

Piezoelectric (PE) and triboelectric (TE) sensing mechanisms are very similar phenomena of mechanical-to-electrical energy conversion. In this subsection, we present the recent progress on stretchable mechanical strain/pressure sensors using PE effect. The PE effect is shown in Figure 3a and is based on the ability of materials to generate electrical charges under external mechanical force, pressure, or strain.[88–95] When a tensile or compressive force is applied over the top of the PE device (Figure 3a), piezopotential difference, between the top and bottom electrodes, is generated. The essence of the potential developed is the relative displacement of the cations and anions centres in the PE material, resulting in a microscale dipole moment. Polarization from all of the units, inside the material, results in a macroscopic potential drop, called "piezopotential," along the straining direction.

Conventional PE sensors have the advantage of being energy autonomous, but are limited for wearable applications due to their intrinsic lack of stretchability. In the literature, two different strategies to achieve stretchability in PE devices have been reported: (i) implementing intrinsically stretchable materials, and (ii) geometrical device engineering. Recent advances in the field of material and nanoscience provide a promising option to develop intrinsically stretchable materials, such as nanomaterials[96], organic/inorganic hybrid networks[95] and so on. Alternatively, in order to maintain the high electrical performance of conventional rigid materials, geometric design engineering has been explored to obtain stretchable PE devices[97].

Many different PE materials, including lead zirconium titanate (PZT)[88,89,91,96], ZnO NWs[92,95,98] and poly(vinylidenefluoride-co-trifluoroethylene) (PVDF-TrFE)[90,93] PVDF[99] and device engineering[97] have been explored to develop wearable PE based strain sensors showing great potential for real-time recording of physiological conditions and body motions. For instance, Dahiya et al.[95], have developed a highly sensitive tactile sensor using PE organic/inorganic hybrid materials. The device sensitivity is demonstrated by wearing it onto the skin of a human index finger. A gentle strain induced by the bending of the finger generated a peak open-circuit voltage of more than 3 V. On the other hand, Sun et al.[97] have exploited geometric design engineering approach to realize PE sensors as wearables. They report an integrated stretchable PE sensing system in conjunction with wireless electronics for continuous health monitoring. This device is composed of two subsystems, a kirigami-based stretchable and self-powered sensing component, and a wireless communication interface for data transmission. The effectiveness of this approach for implantable devices is demonstrated by measuring the surface strain of a deforming balloon and *ex vivo* pig heart, and as a wearable sensor by measuring knee flexion.

### 2.1.4 Triboelectric sensors

TriboElectric NanoGenerators (TENGs), introduced in 2012 by Z. L. Wang group at Georgia Tech USA[100] are a promising technology to convert mechanical energy to electrical energy because of exciting device features such as self-powered, easy to fabricate and high performance. The device operational mechanism is based on coupled triboelectrification and electrostatic induction phenomenon and is schematically shown in Figure 3d. Since 2012, significant progress has been made in TENGs by investigating new functional materials and manufacturing techniques. Much of the efforts are made to realize wearable TENGs, which may play an important role in sensing physiological vital signs[101]. These ultra-flexible and light skin-like devices have the potential to be a power supply or sensor element, to meet the requirements for wearable electronics. In this section, some of the best demonstrated TENG devices for obtaining physiological signals are presented. TENGs as potential power source will be discussed in the later sections of the article.

To enhance the performance of the TENGs, two main factors have been explored, 1) choosing materials from different extremes in the TE series (one electropositive and another electronegative) and 2) modification of surface morphologies to enhance the contact area and TE effect. TENGs with nanopatterns have been realized to biomimick the nature[102]. This is achieved using either nanolithography and etching techniques such as electron-beam lithography and/or block copolymer lithography (BCL), and/or dry and wet etching techniques[103–106]. Physiological signals can be obtained from the electrical signals generated by TENGs such as heart rate,[107] pulse,[104] cardiac arrhythmias, and respiratory rates[103]. Ouyang et al.,[104] have demonstrated self-powered pulse sensor (SUPS) with high sensitivity. The nanostructures are obtained using inductively coupled plasma (ICP) etching process. This TENG-based mechanical sensor works with the vertical contact–separation mode. The SUPS when worn over radial arteria can generate output voltage, current and transferred charges of 1.52 V, 5.4 nA, and 1.08 nC, respectively. Both PE and TE based strain sensors have the advantages of fast response, high sensitivity, negligible power consumption but limited stretchability and an issue to be resolved[108]. Both sensor types are more suitable for autonomous detection of dynamic pressure with wide frequency range[109].

### 2. 2 Biochemical sensors

The advancement in the field of nanotechnology has enabled the realization of miniaturized and cost-effective biological and chemical (biochemical) sensors for health monitoring. Wearable BioChemical Sensors (BCS) are considered as the next generation analytical methods for an emerging third wave of technology to replace bulky and expensive analytical instruments in the health care industry[2,119–121]. In health monitoring, it is important to detect the presence of analytes into the body´s biofluids such as sweat using wearable biochemical sensors such as: ions (sodium, calcium, potassium, chlorine, pH), glucose, lactate, enzyme (alanine and aspartate aminotransferase, tyrosinase), proteins (troponin, C-Reactive Protein, BNP), hormone (cortisol), alcohol, drugs and antibodies[29,120,122–125]. However, the correlation between the level of analytes in sweat and the health condition has not been totally understood.

Wearable BCS are devices that can adhere to the body skin and designed in a way to integrate special receptors to detect the presence of biomarkers in human´s biofluids that can be either interstitial fluid (ISF) or sweat. Wearable BCS devices have additional and indispensable requirements compared to classical biochemical sensors such as: (i) sensitivity to analyte in low concentrations between fM to mM, (ii) stretchability and mechanical stability to ensure a good adhesion to the skin (iii) flexibility to certain extent to allow a comfortable movement, (iv) response and recovery times that are safe in case of no drift, (v) low power consumption for continuous monitoring, and (vi) overall sensor lifetime. The first and the last properties are much difficult to fulfil but possibility to replace the sensor might be an adequate solution.

**2.2.1 Invasive BCS**: The access of wearable biochemical sensors to interstitial fluids (ISF) from the skin is usually invasive because they necessitate microneedles to penetrate the epidermal/dermal tissues and ensure the transport of the ISF to biomarkers´ receptors on the wearable BCS to skin. Nevertheless, the access to ISF is invasive, which represent a concern and needs the development of minimally invasive microneedles[126]. In this category of minimal invasive wearable, the glucose sensor is the most well-known commercial device as an easy-to-use sensor indicating the sugar level of diabetic patients with high precision.

**2.2.2 Non-invasive BCS:** Alternatively, a non-invasive wearable BCS device can use sweat biofluids to detect different analytes. In contrast to ISF, sweat is produced on the skin naturally without invasive methods[2]. From sweat, the wearable BCS can quantify various biomarkers. In literature, several human biomarkers are detected into sweat such as sodium, chlorine, potassium, lactate, calcium, glucose, ammonia, ethanol, urea, cortisol, and various neuropeptides and cytokines.[5] Table 3 summarizes some of the best reported work on wearable BCS sweat sensors. Various different sensing mechanisms are used to measure analytes in sweat such as colourimetric, optical, electrochemical and impedance-based sensing. The most common and versatile method is electrochemical detection, which measures electrical potential or current of transduced analyte concentrations[5].

**2.2.2.1 Detecting Ions**: Most wearable sensors based on electrochemical detection of ions ($H_3O^+$, $Na^+$, $Ca^{2+}$, $K^+$, $Cl^-$) in sweat are based on potentiometric measurements using ion-sensitive membranes specific for each ion[127–130]. For instance, Bandodkar et al.[127] have developed a tattoo-based potentiometric ion-selective sensor for epidermal pH monitoring through screen printing fabrication approach. The fabricated tattoo sensor encompasses a rapid and near-instantaneous response to pH of human sweat in the range 3–7 yielding 80% of their steady-state signal within the first 10 s while a completely stabilized signal is observed within 25 s. The applied deformation to the tattoos improves the sensor response to 59.6 mV/pH unit within the first 10 bending iterations. Electrochemical method is also applied for testing cystic fibrosis (CF) disease on individuals by sensing $Cl^-$ concentration in sweat[131]. To use a standalone device, complex and bulky electronic circuits with power-on board are required. For this reason, Ortega et al.[132] conceived a self-powered smart patch for sweat conductivity monitoring as a screening test for CF. The patch consists of a paper battery that is activated upon absorption of sweat. The developed battery consists of two coplanar electrodes: anode, made of magnesium, and a cathode, made of silver chloride. The electrodes are positioned at 1.5 mm distance and covered with two layers of glass fiber-based paper, which has a high porosity and water absorption rate. By connecting a purely resistive load of 2 kΩ, the voltage generated by the battery is directly proportional to the sweat conductivity that is absorbed by the electrolyte (paper layer) of the battery.

**2.2.2.2 Detecting Glucose and lactate**: Glucose and lactate levels in bodily fluids are usually measured by amperometric measurements where enzyme-based sensors produce electric current caused by a redox reaction. Redox reactions in glucose oxidase (GOx) or lactate oxidase lead to a decrease in oxygen concentration and liberation of hydrogen peroxide ($H_2O_2$), which is detected by the sensor and directly proportional to the glucose or lactate concentration. For glucose, several GOx-based sensors are developed. In the first generation, GOx-based sensor monitors $O_2$ consumed or $H_2O_2$ produced, which indicate the amount of glucose consumed. Later, the incorporation of platinum nanoparticles (Pt NPs) and Prussian blue (PB) as an electrocatalytic mediator allows to increase the sensitivity of glucose detection by favoring a selective oxidation of $H_2O_2$ and lowering the reaction potential to near 0 V versus Ag/AgCl[133–135]. In the second generation, redox mediators as ferrocene are added to interact with enzyme, but such material could not be used in vivo due to its toxicity. The third generation GOx-based sensors use engineered enzymes to facilitate the direct electron exchange between electrodes and enzymes. For example, nanostructured electrodes, such as electrodes treated with CNTs, could be coupled to GOx but reaction still need $O_2$ supply. In addition to glucose oxidase, glucose dehydrogenases have the advantage to induce a redox reaction independently of $O_2$ concentration but have the disadvantage of being not specific to glucose (maltose) and thus, could overestimate the measured glucose concentration[136].

Lactate is a biomarker component reflecting a person´s physiological condition. In this context, Kudo et al.[137] developed an enzyme based Lactic Acid biosensor that controls the sweat flow and continuously measures the $H_2O_2$ resulting from the enzymatic reaction of Lactate with Lactate oxidase. The biosensor consists of two working electrodes and a counter electrode made of carbon graphite and a reference electrode Ag/AgCl. A three-dimensional flow channel is bonded to the transducer for collecting sweat by means of a sucking pump. A sensitivity of 0.16 nA $\mu M^{-1}$ is obtained if the Lactate concertation varies between 10 µM and 1.0 mM. Neethiraja et al.[138] explored the functionalization of anti-lactate antibody onto electroreduced 2D graphene sheets for the fabrication of an electrochemical immunoassay system for Lactate concentration ranging between 0.15 and 25 mM with a detection limit of 0.1 mM. A potentiometric lactate sensor based on the incorporation of lactate oxidase into a hydrogel matrix is demonstrated by Tokito et al.[139]. The measured potential is linear for a concentration ranging between 0 and 1.2 mM. The steady-state potential difference after injection of Lactate is dependent on the load resistance between the electrodes as well as on the hydrogel (agarose gel) thickness.

**2.2.2.3 Detecting Hormone – Cortisol:** Cortisol also called stress hormone, is a steroid hormone secreted by the hypothalamic-pituitary-adrenal system. It is a well-known biomarker for regulating metabolism, immune response and circadian rhythm as well as leveling the psychological stress, that´s why it is called a 'stress hormone'[140]. The current strategies for measuring unbound cortisol are limited to laboratory techniques such as chromatographic techniques, which are a complex system requiring multistep extraction/purification. For point of care methods, the most developed work for detecting cortisol are based on electrochemical immunosensing method[141,142]. An immuno-complex formation is realized for a sensitive detection of Cortisol using Anti-Cortisol antibodies covalently immobilized on sensing electrodes,[142] polymer composite, ionic liquid polymers;[143] nanoparticles[143,144] and 2D nanomaterials[145].

**2.2.2.4 Detecting Enzyme – Tyrosinase:** Melanoma is an aggressive skin cancer; an early diagnosis of this disease can lead to a reduction in mortality. Currently, the identification of melanoma involves time-consuming complex techniques, aimed at detecting melanoma biomarkers, such as melanocytic lineage markers. Recently, Ciui et al.[146] developed for the first time an electrochemical sensor for detecting enzyme biomarkers on the skin for cancer screening applications to avoid biopsies, that necessitate an invasive method. The designed wearable bandage with and without microneedles for electrochemical sensing allows to detect tyrosinase enzyme (TYR) both on and under the skin´s surface. Tyrosinase is a polyphenol oxidase involved in the synthesis of melanin, which is indicative of an aggressive skin cancer melanoma. The transducer is a three-electrode system on a flexible substrate, that consists of screen-printed Ag/AgCl and Carbon electrodes. The sensing layer consists of a catechol embedded into agarose gel. In the presence of TYR in the skin, TYR will oxidize catechol to BQ quinone product, which generates an amperometric current under −0.25 V potential applied to the electrode. The generated current increases linearly with the amount of tyrosinase enzyme present in the skin and ranging from 0.1 to 0.5 mg $mL^{-1}$.

We believe that a new generation of biochemical sweat sensors should be developed that are noninvasive and self-powered. There are many challenges to be addressed, such as the amount of sweat should be controlled in a precise way to reach a high sensitivity. However, controlling the sweat amount necessitates an additional power source to pump precisely the sweat to the vesicles and on the surface of the functionalized transducer. Moreover, new materials are needed that have soft properties to adhere to the skin with sucking properties to extracted sweat from human skin and transport a define amount to the receptors of the electrochemical transducer. Hydrogels are suitable materials to design innovative biocompatible wearable sweat sensors without damaging human tissues and replace fluidic devices that require several technological realization steps. The integration of the 2D materials with different pore diameters to the sensing layer of the BCS might allow to filter and sense properly the ions in sweat with an unprecedent sensitivity and selectivity. The unique properties of 2D material-based electrodes for electrochemical process offer new possibilities to address the detection challenges of sweat´s biomarkers in low concentration and replace in the future the invasive blood sampling. Therefore, intensive simulation and experimental research works are undergoing to investigate 2D materials advantages for electrochemical sensors.

### 3. Flexible energy storage (Batteries and Supercapacitors)

Battery is an important element of wearables. Existing commercial lithium-ion batteries are heavy and can take lots of space. While the amount of active materials should be limited as much as possible, the battery capacity is directly related to their total amount. While component miniaturisation has made tremendous progress, comparatively little progress has been made on the miniaturization of batteries which has become the key challenge of wearables. If the battery thickness is increased ,the distance of lithium-ions transport also increases while out-of-plane electronic conductivity decreases, resulting in power density decrease and causing poor performance. These limitations are particularly important for microbatteries with the volume of 1 – 10 $mm^3$ and typically used in micro-electromechanical systems (MEMS), micro sensors, micro actuators and micro-biomedical devices. Evidently, the volume of the battery limits the miniaturization of the whole system, and of course, limits the form factor of the wearable sensor as a whole. Based on form factor, energy storage devices are classified into two categories: (i) flexible, and (ii) stretchable energy storage devices.

**3.1 Flexible energy storage devices**: The flexible nature of werables requires a lightweight power source with stable output under various deformation conditions. Additional limitations specific to batteries are material selection, cell chemistry, design and safety[150]. Constructing a flexible energy storage system (i.e. batteries and capacitors) requires conducting substrates, charge storage materials and an electrolyte system. These components are generally arranged in layered,[151–154] planar[155–158] or cable/fiber[159,160] shaped configurations as shown in Figure 4.

**3.1.1 Layered configuration**: The layered configuration or sandwiched design is the most convenient and common method of constructing an energy storage system where electrolyte is sandwiched between two facing electrodes (Figure 4a). This configuration allows for optimal ionic diffusion while mechanical flexibility is achieved by replacing heavy and rigid current collectors with carbon-based substrates such as graphene and CNTs[161]. Such substrates provide large surface area for high mass loading, thus enhancing the charge storage.

**3.1.2 Planar configuration**: In planar configuration, the electrodes are placed side by side while the electrolyte is place on top of electrodes filling the empty space (Figure 4b). Due to its planar geometry, different fabrication techniques like inkjet printing,[162] screen printing[163] and roll to roll printing[164] can be employed to fabricate electrodes with high precision. Width of each electrode and the gap between them defines the electrochemical performance of these energy storage systems[165]. In a recent study, Sun et al.[166] investigated the effect of electrode thickness by printing multiple layers of electrode material and found that volumetric capacitance of a supercapacitor linearly increases with the number of printing layers. However, printing compact designs with shorter interspaces and thicker electrode layers is challenging due to the rheological properties of the conductive inks which consists of binders, solvents, additives and active materials.[167] Moreover, conductive inks formulations for supercapacitor fabrication are relatively simple compared to batteries which incorporate electrode materials such as Si, lithium titanate, $Li_2TiO_3$ (LTO), lithium cobalt oxide, $LiCoO_2$ (LCO), lithium manganese oxide, $LiMnO_2$ (LMO) and lithium nickel cobalt aluminium, (NCA) making it difficult to print at high precision.[155] Nevertheless, various strategies including layer-by-layer printing[166] have been developed to fabricate micro supercapacitors as well as lithium-ion and zinc air batteries in planer and layered configurations (Table 4).

**3.1.3 Fiber-based configuration**: Another novel architecture of flexible energy storage system is to directly use fiber-based electrodes arranged in parallel, intertwined or core-shell configurations (Figure 4c). Unlike planar or layered design, the cable-based electrodes can be woven on textiles in various shapes. Two main requirements for such design are flexible and conductive fibers. An obvious choice is to incorporate metallic wire; however, flexibility and weight of metallic wires limit their use for wider applications. In contrast, carbon-based fibers have become a preferred choice for a light and flexible cable-based energy storage device providing large surface area for charge storing active material. For example, aligned multiwall carbon nanotubes fibers coated with gel electrolyte have been twisted to fabricate a thin supercapacitor.[169] Graphene fibers have also been successfully tested as fiber electrodes in both parallel and coaxial configuration (Table 5) [170,171]. Moreover, high conductivity of CNT fibers and catalytic activity of graphene fibers have been utilized to fabricate CNT/graphene composite fibers[172]. Composite pairing in a heat shrinkable tube is a dominant approach for stretchable battery systems, and have been demonstrated to retain 84% specific capacity after 200 cycles with up to 100% elastic stretching[174]. Figure 5 summarizes the various structures for flexible and stretchable Li-ion batteries.

**3.2 Stretchable energy storage devices:** Although this research is nascent, many advances have been made by introducing new materials and optimizing structures for high electrochemical performance suited to wearable devices. Some of the important and interesting design strategies for wearable-compatible soft batteries are discussed in the following section.

**3.2.1 Structural engineering**: Fiber-shaped LIBs are of interest as exhibit many unique advantages such as flexibility, weavability, and wearability. The last point has been addressed more recently, by engineering alternative form factors and electrode designs to enable unique human-compatible wearability of Li-ion and alternative batteries. The two most common are wavy-type structures, and serpentine or island-bridge structures that take inspiration from elastomer-based skin electronics (Figure 6a-d) and body-compatible wearable sensors. In wavy-type approaches,[176] stretchability without breakage or tearing is made possible by imprinting waviness into the substrate. Such devices can be stretched up to 300%, while maintaining capacity densities of ~1.1 mAh $cm^{-2}$.

**3.2.2 3D printed energy storage devices**: 3D printing rechargeable batteries open up the possibility to design power solutions that are seamlessly integrated into the form factor of wearables, medical devices, consumer electronics, IoT sensors, electronic peripherals and more. The flexible form factor of such devices is realized as sprayable or paintable batteries,[177] and stretchable/flexible battery systems,[176,178]. However, their energy density is limited for wearables and sensor applications. Ultrathin batteries benefit from standard Li-ion cell manufacturing process, with limited capacity the primary trade-off for their svelte profile. 3D printing based on fused deposition modelling (FDM) has been the most accessible and most common method to develop poly(lactic acid) containing graphite, or ABS-type plastics to print battery cells, or to accommodate active materials for electrochemically active battery electrodes.[179] First developments targeted 3D printed microbatteries[180] and composite batteries[181], where recently efforts were undertaken to realize the 3D printable Li-ion battery using a fused filament fabrication (FFF) or fused deposition modelling (FDM) (Figure 6e-g). Since then, numerous iterations of this approach have been reported, including for example, MXene materials incorporated directly into FDM printed plastic for current collector-free supercapacitor applications with a capacitance retention of 90% over 10000 cycles at a rate of 0.2 A $g^{-1}$.[182] The search for an optimized route to 3D print all components of a battery, either separately or sequentially (anode, separator, cathode and outer casing) is underway[183,184]. Reyes et al.[185] and Maurel et al.[186] have made recent advances in controlling the additives and conductivity of PLA, making a 3D printed cell and enhanced conductivity PLA anode. Low electronic conductivity was one reason for modifying the graphitic content of PLA, as identified by Foster et al.[187]

The compact nature of flexible energy storage systems makes them susceptible to frequent charging. Therefore, developing a micro-integration system for energy generation and storage is vital for their practical usage in autonomous wearable sensors for the development of EAWS device system. An effective strategy is to design a flexible electrode, which simultaneously serves for energy harvesting and storage. For instance, in a recent attempt, Peng et al.[195] fabricated fiber based electrodes to construct a wire shaped dual function energy storage device. The two fiber electrodes consist of $TiO_2$ modified Ti wire and aligned CNT fiber, which were twisted together inside an Iodide based electrolyte. Under illumination, the device functioned as a dye sensitized solar cell (DSSC) with 6.58% conversion efficiency as well as electrical double layer capacitor with a specific capacitance of 85 μF $cm^{-1}$. In another design, the same group used a core-sheath structure with photoelectric converter as sheath and lithium ion battery as a core fiber.[163] The inner core consists of LMO/CNT and LTO/CNT fibers, while CNT layer was used as an outer energy harvesting part providing photovoltage of 5.12 V. Beside solar harvesting, a novel integration solution is to use TENG that converts mechanical energy into electrical form. As an example, Dong et al.[196] reported a self-charging yarn-based textile with a fiber based TENG and a supercapacitor. The energy harvesting system was prepared by coating silicon rubber onto the surface of stainless steel and polyester fiber. Upon contact, electrification occurs at the interface of silicon dielectric layer causing a flow of electrical charge. Regulated by a rectifier, the energy harvesting yarn was connected to a yarn-based supercapacitor providing maximal power density of 85 mW.$m^{-2}$. Despite significant advancement, these integration devices need further improvement in terms of aesthetics and wearability, which is the major concern in flexible electronic devices. Recent advancement in stretchable energy harvesting technologies is discussed in the following section.

## 4. Energy harvesting technologies

Energy harvesting is a key component for EAWS to ensure their energy autonomy over long period of time. The four major energy harvesting technologies currently being explored to power wearable sensors are: solar cells, piezoelectric nanogenerator (PENG), triboelectric nanogenerator (TENG), and thermoelectric generator (TEG), as shown in Figure 7. They have their own advantages and limitations in terms of the average generated power and ease of integration onto flexible substrates.

**4.1 Solar cells**: Among the popular alternative energy sources, solar energy is inexhaustible and it is also a form of clean energy. The biggest advantage of using solar cells as the power source is their ease of integration, lack of harmful emission and readily available resources.

For instance, Núñez et al. come up with an innovative device structure consisting of a transparent tactile sensitive layer based on single-layer graphene, and a photovoltaic cell underneath as a building block for energy-autonomous, flexible, and tactile electronic-skin.[202] In an another report, Pan et al.[199] have demonstrated a new and general method to produce wearable DSSC textiles with a high energy conversion efficiency that is well maintained under bending. Outdoor photovoltaic modules could provide sufficient power output for energy intensive applications but it varies with lighting conditions. To mitigate such an issue, indoor photovoltaics have garnered lot of research interest in powering wearable electronic devices. The indoor ambient light composed of natural light coming from outside, and artificial light provided by lamps or luminaries have typical irradiance ranges between 0.1–1 $mW/cm^2$ (depending on the distance between the light source and sensor).[203,204] Among many different types of solar cells investigated, amorphous silicon based or dye-sensitized solar cells have advantage of higher absorption in the visible wavelength range. The harvested power densities using these solar cells range from 37.9 to 92.6 $\mu W/cm^2$ for bright indoor conditions.[205] However, it is important to note that for both outdoor and indoor cases, the peak instantaneous power harvested is not assured at all times because of conditions such as cloud cover, location of sensor, night time or indoor lights turning off. Subsequently, the power demanded by a wearable sensor cannot be met. For example, health monitoring medical patches are generally located under patient's clothes where the amount of light intensity available is very low, making solar cells unsuitable to power such wearable sensors. Thus, to ensure a steady and reliable power supply to wearables, it is necessary to explore additional energy harvesting mechanisms.

In this regard, the ubiquitous kinetic/mechanical energy is a potential source of electrical energy with possibility to fabricate devices over fully flexible and stretchable substrates[97,101]. The natural mechanical energy from human body can be exploited for energy generation using devices called ‘nanogenerators (NGs)’ for the advancement of wearable technologies[206]. For example, both PENGs and TENGs have been used to develop self-powered sensors for monitoring health, and scavenge body movements to power micro-devices.

**4.2 Piezoelectric nanogenerators (PENGs)**: High performance PENGs have been demonstrated with anisotropic PE materials. Among different flexible PENG device prototypes such as laterally integrated NGs (LINGs),[109] fibre-based,[207] and vertically integrated NGs (VINGs),[208,209] VINGs are by far one of the most developed and investigated structures. In the VING device prototype, NWs are vertically grown over flexible substrates and then encapsulated with a polymer, thus forming a hybrid organic/inorganic composite system. Using such a device prototype, Zhu et al.[208] reached a new milestone in the performance of PENG devices, demonstrating peak open-circuit voltage and short-circuit current to a record high level of 58 V and 134 μA, respectively, with a maximum power density of 0.78 W $cm^{-3}$. Although VING devices showed promising energy harvesting performance, some of the key issues restricting the device performance are yet to be addressed, such as figure of merit evaluation, screening effect mitigation, feasibility of technology transfer to fully stretchable system, and efficient transfer of mechanical and electrical energy to and from NWs, respectively.[95] Recently, adopting a comparatively new PENG device prototype, called NanoComposite Generator (NCG), researchers are trying to improve the stretchability of the PENG devices.[210–214] Generally, the NCGs are composed of PE nanomaterial dispersed in an polymeric matrix, which can also be treated as device substrate. The main advantages of NCG device fabrication process are its mechanical robustness, cost effectiveness, and scalability. The NCG device prototypes have been demonstrated using various perovskite nanomaterials such as barium titanate ($BaTiO_3$),[210,214] zinc stannate ($ZnSnO_3$),[213] and sodium potassium niobate ($(Na_{0.5}K_{0.5})NbO_3$)[212]. A hyperstretchable NCG is demonstrated using lead magnesio niobate-lead titanate (PMN-PT) particles in an ecoflex silicone rubber matrix with very long silver NWs as electrode material.[211] The obtained NCG device generated outstanding performance (4 V and 500 nA) and extraordinary stretchability of more than 200%. However, most of the reported NCGs are assembled using perovskite nanostructures that needed high growth temperature (around 1000 °C), extra poling step to align their dipoles in one direction, and conductive additives such as CNTs, which add cost and complexity in the device process. To resolve these issues, Dahiya et al.[95] developed a facile, cost-effective, and industrially scalable process flow for the fabrication of high performance, mechanically robust, stretchable nanogenerator (SNG) on PDMS substrate. The active PE nanomaterial, vertically aligned ZnO NWs, is directly grown on PDMS using low-temperature hydrothermal growth process. An inorganic/organic composite type piezoelectric energy harvesting device is realized by encapsulating the ZnO NWs in a parylene C polymer matrix. The SNG devices exhibit excellent performances with a high open-circuit voltage ≈10 V, short-circuit current density ≈0.11 μA $cm^{-2}$, and peak power ≈3 μW under a vertical compressive force using a mechanical shaker.

**4.3 Triboelectric Nanogenerators (TENGs)**: Likewise PENG, TENG has gained lot of attention recently due to their low cost, easy fabrication, simple structures and high conversion efficiency. The performance of TENGs is based on the effective contact area. Consequently, the morphology of the contact surfaces and architecture of TENGs have been thoroughly investigated to enhance the output performances. Yi et al.[200] developed a soft, stretchable, and fully enclosed self-charging power system by seamlessly combining a stretchable TENG with stretchable supercapacitors. The wearable TENG can transduce diverse mechanical motions into electricity, which is stored in supercapacitors. The developed stretchable power system can be worn on the human body to effectively scavenge energy from various kinds of human motion, and it is demonstrated that the power source is able to drive an electronic watch. Lai et al.[215] reported a skin-like TENG with the advantages of durability and stretchability to power wearable electronics. The developed TENG is fully conformal on various nonplanar or irregular objects, including human bodies, spheres, and tubes, etc., to act as power sources. The open circuit voltage and transferred charge density of (2.5 × 2.5 $cm^2$) TENG device can reach 70 V and 100 $\mu C/m^2$, respectively, under 10 N force. They further demonstrate the first fully self-sufficient and adaptive e-skin system that can map touch by responding with visual light-emitting diode (LED) signals without the need of external power supply.

Both PENG and TENG based energy harvesting technologies are very promising as they can be constructed over flexible and stretchable platform with high performance. However, both device types give high output voltages under open-circuit condition (very high resistive load). Such output voltage decreases dramatically with the decrease of the external load resistance that means additional circuitry is needed for impedance matching. Such an additional circuit will surely reduce the battery life and impose challenges for the development of EAWS system.

**4.4 Thermoelectric generators (TEG)**: Thermoelectric generators (TEG) have also drawn a lot of interest on using body temperature gradient as a source to generate power. TEG generate electric potential using the principle of Seebeck effect. The voltage developed by thermoelectric material is directly proportional to the temperature gradient that different part of the material is subjected to. The following equation explains this dependence mathematically:

$$V = \alpha \Delta T$$

Where, α is the Seebeck coefficient (V $K^{-1}$) of the thermoelectric material and ΔT (K) is the temperature gradient induced across the thermoelectric material. The figure of merit of the thermoelectric material (zT) depends on the Seebeck coefficient (α) of the material, the electrical conductivity (σ), the thermal conductivity of the material (k) and the absolute temperature (T) and is shown by the following equation:

$$ZT = \frac{\alpha^2 \sigma}{k} \cdot T$$

The human body is a continuous source of thermal energy, which nowadays is widely explored for supplying a substantial temperature gradient to the TEG aiding in generation of uninterrupted energy to replace the bulky energy-deficient batteries. The temperature of the human body changes with the physical activity level and the surrounding atmosphere through its metabolic function[216,217]. The body heat is transferred through tissues and blood to skin, which is then dissipated to our surrounding atmosphere by conduction, convection and radiation. The difference in temperature of the human body and the outer atmosphere can be of great use to produce useable source of energy, which can power various wireless transceivers, chemical sensors, ECG sensors by which one can successfully develop a health monitoring system[218–222]. Bismuth telluride based materials dominate with higher thermoelectric properties in wearable applications and fair amount of reviews and works have been extensively reported in the literature[218,223–225].

The first commercially available µTEG, to power a wristwatch, was developed by Japanese company Seiko[226]. The harvested energy of the µTEG was used to power the watch as well as to charge the battery, as the maximum estimated power was 22.5 µW. The next significant commercial development in this field is demonstrated by MicroPelt. They developed their devices using sputtered BiTe or SiGe based thermoelectric materials and using flip-chip bonding approach[227]. The p- and n-type thermoelectric materials were deposited separately on two different wafers and bonded during the final assembly. This approach of the device fabrication offered them the ease of independent material deposition and optimization. However, the main challenge was the bonding step of the device, which requires precise control over the thermoelectric material thickness, bonding temperatures, and the bonding material. For wearable application, in order to exploit the body heat gradient completely, the area of contact between the device and skin has to be sufficient. In this situation, flexible thermoelectric generators offer the tunability of device performance by altering the area of interaction.

Wang et. al.[201] developed flexible thermoelectric device on a flexible printed circuit board with 52 thermopiles for wearable applications. P- and n-type Bi-Te based powder materials is used in the development of the device. Using such a device, an open-circuit voltage of 37.2 mV with peak power density of 16.8 μW/$cm^2$ at a temperature gradient of 50K has been achieved. A DC-DC converter is used to boost the generated voltage to power 3-axis miniaturized accelerometer for detection of human body motion. Hyland et al. designed and developed optimum TEG design and evaluated the performance for powering the electrocardiogram sensors by TEG and tested on the hot-plate as well on different human body locations[228]. The effect of heat spreader, airflow conditions and device position over human body were evaluated. Device positioned over chest and the wrist produced less power compared to the upper arm. The least power was obtained when the TEG device was placed on the T-shirt. TEG placed in all the locations were evaluated with different airflow velocity. Leonov et al. have studied the performance of commercially available TEG integrated in clothing for powering wearable electronics and to obtain self-powered wireless sensors nodes using body heat.[216,229] In an another study, Lay-Ekuakille et al. have shown that TEG able to extract warmth from body tissue to supply power to hearing aids for deaf patients.[230] An off-the-shelf TEG with an area of 3×3 $cm^2$ is employed which generated power of 20 μW at 22 °C corresponds to a power density of 2.2 μW/$cm^2$.[231] This device produced a maximum of 12 mV across a suitable load at the point of maximum power transfer. Recently, self-powered wearable ECG using body heat through TEG for medical application has been demonstrated[232]. Owing to the usage of a polymer-based flexible heat sink, the maximum temperature difference was obtained. With this setup, a maximum power density of 38 μW/$cm^2$ was recorded which was enough to drive the power management circuit along with the ECG system. Recently, Lal et al.[233] have demonstrated the fabrication of flip-chip bonded cross-plane configured μTEG device using electrodeposition technique on silicon wafer. The effective device size was 5×5×1.2 $mm^3$ consisting of 210 effective thermoelectric pairs of Bi-Te and CuTe pillars as p- and n-type, respectively. An open circuit voltage of about 315 mV was measured for a temperature difference of 10 K using a complete device with 210 TE pairs. These innovative works on stretchable energy harvesters will lead to promising improvements in future autonomous wearable electronics.

## 5. Challenges and future outlook

Since the launch of first commercial wearable sensor in 1960s, wearable devices have recently begun to improve healthcare services. Such an improvement can be dedicated to the miniaturization of sensing devices, and the rapid progress in microelectronics and wireless communication technologies. Some of these modern-day wearable sensors are dominated by commercial wrist-watch sensors such as Apple Watch[10], and medical patches such as the iRhythm Zio patch[234] and Abbott's FreeStyle patch[235] for continuous health monitoring. However, these wearable sensors have limitations such as inaccuracies, large power consumption, short battery lifetime and no multi-sensing functionality. Therefore, several technical challenges still need to be addressed to develop EAWS for personalized healthcare. Some of these challenges are discussed in this section.

1) **Component integration and wearability**: One of the challenges is the integration of different EAWS components on a single package or substrate. Wearable systems focus on different properties than rigid systems. Properties such as flexibility and stretchability become important because they greatly influence the wearing comfort/convenience. When it comes to the design of a wearable system, the mechanical properties of every component become important. The flexibility of components varies between flexible, semi-flexible and rigid.

Components have to be placed on a flexible substrate in a way so that the system itself remains flexible. Same applies to the stretchability of a system. Furthermore, because of the variety of different electronic components, energy harvesters, sensing units, an energy processing unit, a processing and communication system, the components come in different sizes and packages, thus, different assembly techniques need to be used such as conductive adhesion, soldering or bonding. Therefore, integration of various device components on a single flexible and stretchable package is a major concern for EAWS system.

**2) Power consumption and to find suitable source of energy**: Another challenge is to find efficient strategies to supply power for continuous operation of EAWS system. A smart system is continually collecting and analyzing data by using a microprocessor, or communicating to the cloud. Regardless of communication approach, EAWS require continuous power supply to power sensors and electronics. Currently, supplying electrical power to wearable sensors is done by using conventional rigid lithium batteries. Apart from their large size and weight, such batteries require frequent replacements. Alternative approaches for energy harvesting are being investigated[101,176,178,232,236–238] where electrical energy is harnessed from the ambient (solar cells, thermoelectric, piezoelectric, triboelectric or combination of sources through hybrid generators) or kinetics (body dynamic movements) that can be stored in tandem with a rechargeable and stretchable batteries. However, the incorporation of energy harvesters comes with new challenges regarding the system design such as system integration and rechargeability.

**3) Miniaturization**: Wearable systems need to be comfortable for the user and therefore, the goal is to make them as small and thin as possible without sacrificing functionality. The trade-off between comfort and energy harvesting capabilities depends on the application, thus miniaturization of the device is another issue to be resolved. 3D integration technology offers major advantages in all of these aspects and definitely will be needed in the mid-term horizon frame for miniaturized high-performance medical devices. Nevertheless, to accommodate the limited power capacity of a wearable system, several low energy communication protocols are available such as BLE or ZigBee and are supported by several microcontrollers. A common approach for wearable healthcare systems is to collect data, do some minor data processing and store the data temporarily. Periodically the stored data is sent and processed in a central system. If the system senses an immediate danger, an emergency signal is sent out. Therefore, adapting and optimizing data sampling frequency and data transmission (according to the stand-by and active mode of the sensor) could increase efficiency of the EAWS system, and thus reduce the power consumption.

**4) Sensor robustness:** Repeatability of sensing performance of the developed wearable sensor in lab-to-fab environment is a concern. Most of the best reported wearable sensors are often prototypes unsuitable for use on patients/individual in day-to-day life conditions such as sleeping, swimming, running, etc. Even the well-designed wearable systems, which have demonstrated a good biosignal sensing in lab testing, have difficulty operating with sufficient robustness in real-life conditions. Thus, the durability and robustness of wearable sensors needs further improvement. The lack of clinical testing of the device results in poor device repeatability, which are prone to movement artefacts, a major challenge in wearable system design[239]. Innovative packaging technologies, which are less susceptible to noise from motion, may help to realize robust wearable sensing system.

**5) Multi-sensing:** The majority of commercial wearable sensors are capable of monitoring single physiological health parameter whereas multi sensing functionality is desirable to monitor several vital signals that can be essential for meaningful home health care, sporting activity and remote patient-doctor communication. Integrating multiple sensor devices would not be an ideal choice due to complexity with assembly and calibration and moreover, uncomfortable for the users. Hence, a multi-modal wearable sensing device that simultaneously monitors heart rate, respiratory flow, temperature, arterial oxygen rate and extracts data from biomarkers would be desirable to provide an overall assessment of the individuals' health.

**6) Human safety**: Safety issues are paramount for any new form factor batteries or power sources for wearables and sensors. This is especially important as sensors are in close contact with human skin. Flammable materials and electrolyte, toxicity and other factors are important consideration when designing new EAWS components. From an integration and deployment point of view, all components will need to undergo regulatory testing for use and transport, as it is currently required for all current power sources and batteries.

**7) Data security**: Last but not least, one of the most significant challenges to deploy EAWS is data security. IoMT technology monitors patient's health status in real-time via sensing and transmitting data through a connected network. The connected network has a significant potential risk of security attacks. Cybercriminals can hack the system and misuse patient's data impose privacy and security challenges. For example, the attacker can misuse patient's data to create fake identity to purchase medical equipment/drugs, which they can sell later and file a fake insurance claim in patient's name. Effective implementation of cyber safety programs for early detection of such online incursion will ensure the confidentiality of the healthcare data generated by EAWS system.

To conclude, the IoMT technology has the potential to bring health management into the modern era, empowering the user, and enabling doctor – patient mobile health monitoring system. The IoMT can help in continuous health monitoring and notify healthcare providers with actual data to identify issues before they become critical. The 'self-powered EAWS' device system has been presented here as one of the important device technologies to enable the vision of IoMT.

**Acknowledgements**

This work was financially supported through SmartVista project, which has received funding from the European Union's Horizon 2020 research and innovation programme under the grant agreement No. 825114.

**Conflict of interest**

The authors declare no conflict of interest.

# Figure captions

**Figure 1.** Illustration of Internet-of-Medical-Things (IoMT) for personal healthcare.

**Figure 2**. Schematics illustrating the sensing mechanisms of: a) Piezoresistive and b) Capacitive. $R_C$ is contact resistance, $R_D$ is device dimensional resistance, $R_T$ is tunnelling resistance between two 1D material, and $R_I$ is intrinsic resistance of the 1D material.

**Figure 3.** Schematics illustrating the sensing mechanisms of: a) Piezoelectric; b) Triboelectric.

**Figure 4**. Schematics for three common designs for flexible energy storage systems. (Adapted from Ref[168] with the permission of Royal society of Chemistry).

**Figure 5.** (a) Structure of the flexible wire-shaped lithium-ion battery composed of aligned woven yarns of active materials. (b) The wire-shaped battery can power a LED when stretched to double its length. (c,d) Fiber-shaped lithium-ion batteries directly woven into flexible textiles. Reproduced with permission from Wiley-VCH, 2014[174] and the American Chemical Society, 2014[175].

**Figure 6.** (a) Schematic of the island-bridge structure design for stretchable batteries with (b) an exploded-view layout of the various layers in the battery structure. (c) Illustration of "self-similar" serpentine geometries used for the interconnects (black: first-level serpentine; yellow: second-level serpentine) (d) Capacity vs cycling for the device. Reproduced with permission form Nature Publishing Group.[188] (e) 3D printing of MXene-type 2D material inks with suitable viscoelastic fingerprints layer by layer for current-collector-free electrodes. Reproduced with permission from Wiley-VCH, 2019.[182] (f) Single print PLA-based Li-ion battery powering an LED and (g) Capacity and Coulombic efficiency of the single print battery at a current density of 20 mA $g^{-1}$ for 10 cycles. Reproduced with permission form American Chemical Society, 2013[185].

**Figure 7.** Types of wearable energy harvesting technologies: Solar cells;[199] PENGs;[97] TENGs;[200] and TEG[201].

# Table captions

**Table 1.** Representative summary of the performances of piezoresistive and capacitive wearable strain/pressure sensors reported. *[SWCNT: Single walled carbon nanotubes; Au: Gold; PVDF: Polyvinylidene fluoride; PU-PEDOT:PSS: poly-urethane (PU) and poly(3,4-ethylenedioxythiophene):poly(styrenesulfonate) (PEDOT:PSS)]. Gauge factor for capacitive strain sensor is given by:* $GF = \Delta C/C_0\varepsilon'$ *where* $\Delta C$ *is change in capacitance,* $C_0$ *is the initial capacitance and* $\varepsilon'$ *is the applied strain.*

**Table 2.** Representative summary of the performances of piezoelectric and triboelectric wearable strain/pressure sensors reported. *[PTFE: Polytetrafluoroethylene; PET: Polyethylene terephthalate; FEP: Fluorinated ethylene propylene]*

**Table 3**. Wearable biochemical sensors towards measuring biomarkers in sweat with operation features of the device that are novel and recently published. *[CNT: carbon nanotubes; PVC: Polyvinyl chloride; PANI: Polyaniline]*

**Table 4**. List of selected flexible batteries with three different designs and cell chemistries. [CC: *Current Collector; NCA: Nickel Cobalt Aluminum Oxide; PAA: Polyacrylic Acid; PEO: Polyethylene Oxide; LCO: Lithium Cobalt Oxide; LMO: Lithium Manganese Oxide; LTO: Lithium Titanate; and LiTFSI: Lithium bis(trifluoromethanesulfonyl)imide.*

**Table 5**. List of selected flexible supercapacitors with three different designs and fabrication methods. *TMOS: Tetramethyl Orthosilicate; FA: Formic Acid; EMI TFSI: 1-ethyl-3-methylimidazolium bis(trifluoromethylsulfonyl)imide; PVA: Polyvinyl Alcohol; PC: Polycarbonate*

**Figure 1**

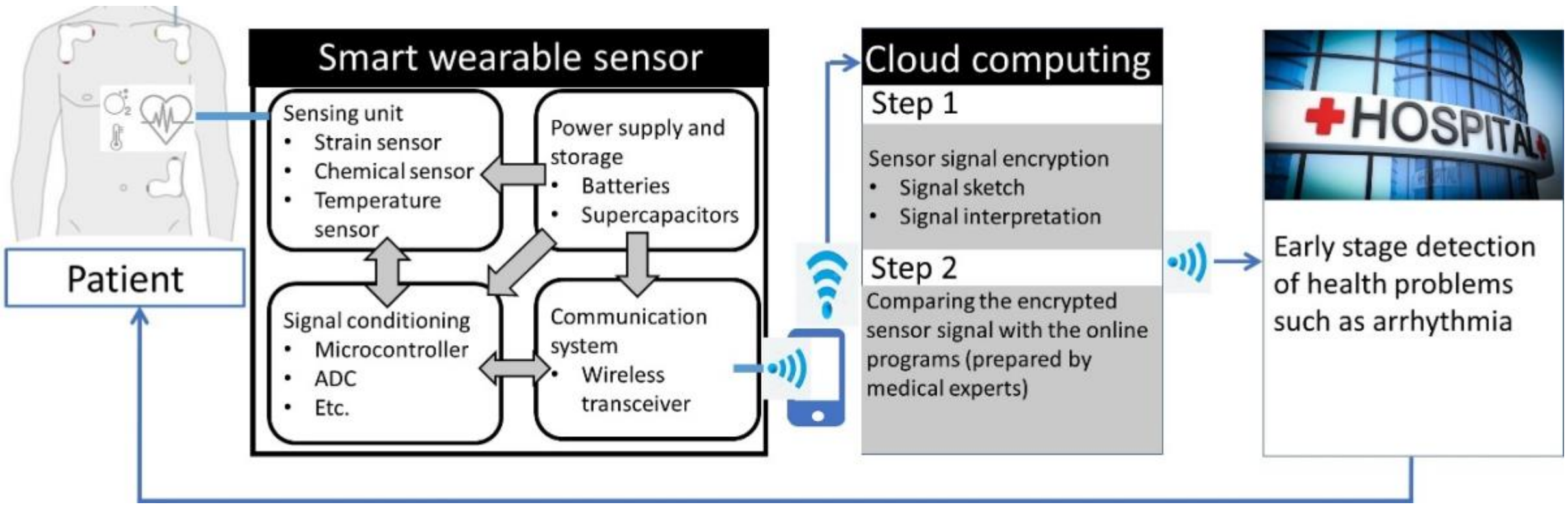
Smart wearable sensor
Sensing unit
• Strain sensor
• Chemical sensor
• Temperature sensor
Power supply and storage
• Batteries
• Supercapacitors
Signal conditioning
• Microcontroller
• ADC
• Etc.
Communication system
• Wireless transceiver
Patient
Cloud computing
Step 1
Sensor signal encryption
• Signal sketch
• Signal interpretation
Step 2
Comparing the encrypted sensor signal with the online programs (prepared by medical experts)
HOSPITAL
Early stage detection of health problems such as arrhythmia

**Figure 2**

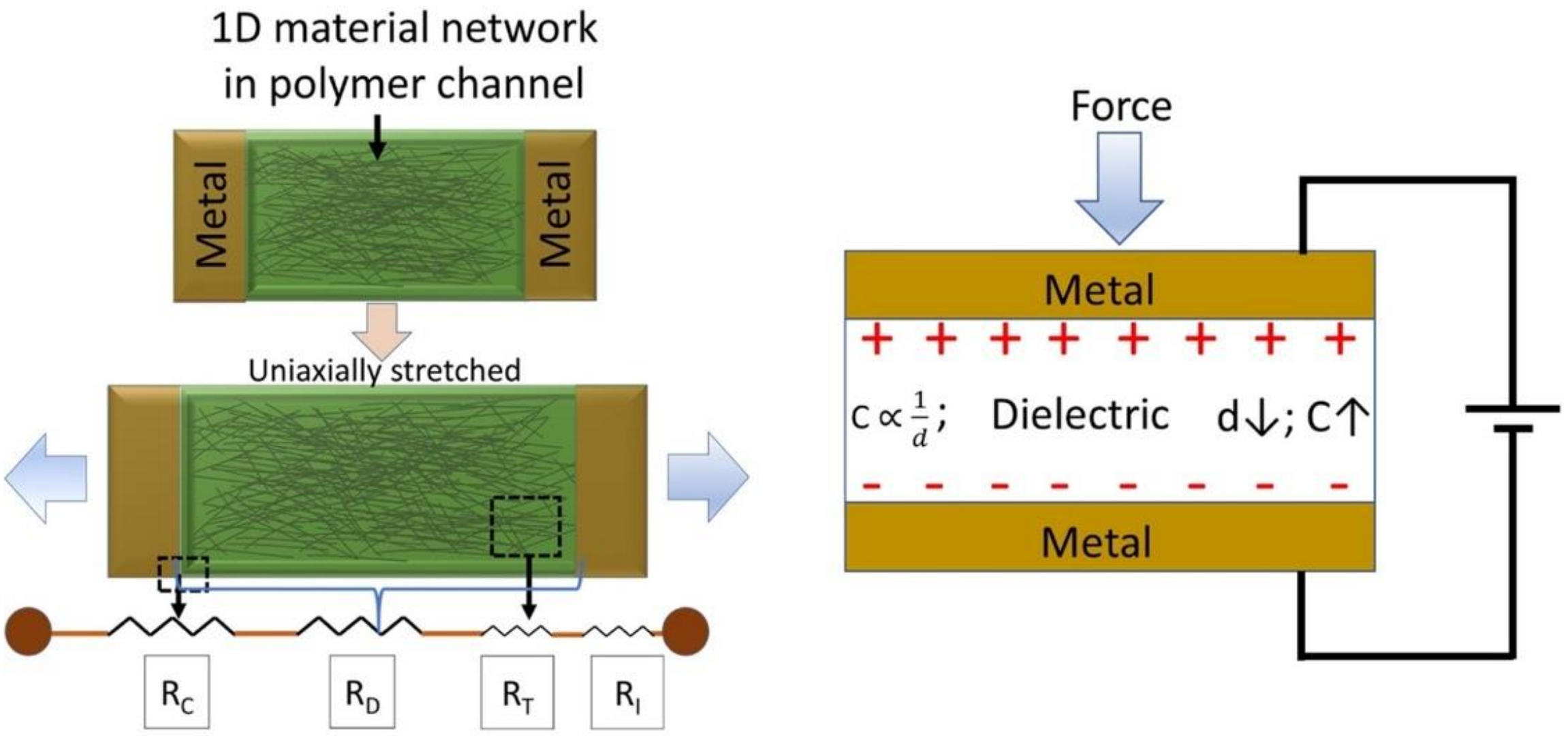

a) Piezoresistive b) Capacitive

**Figure 3**

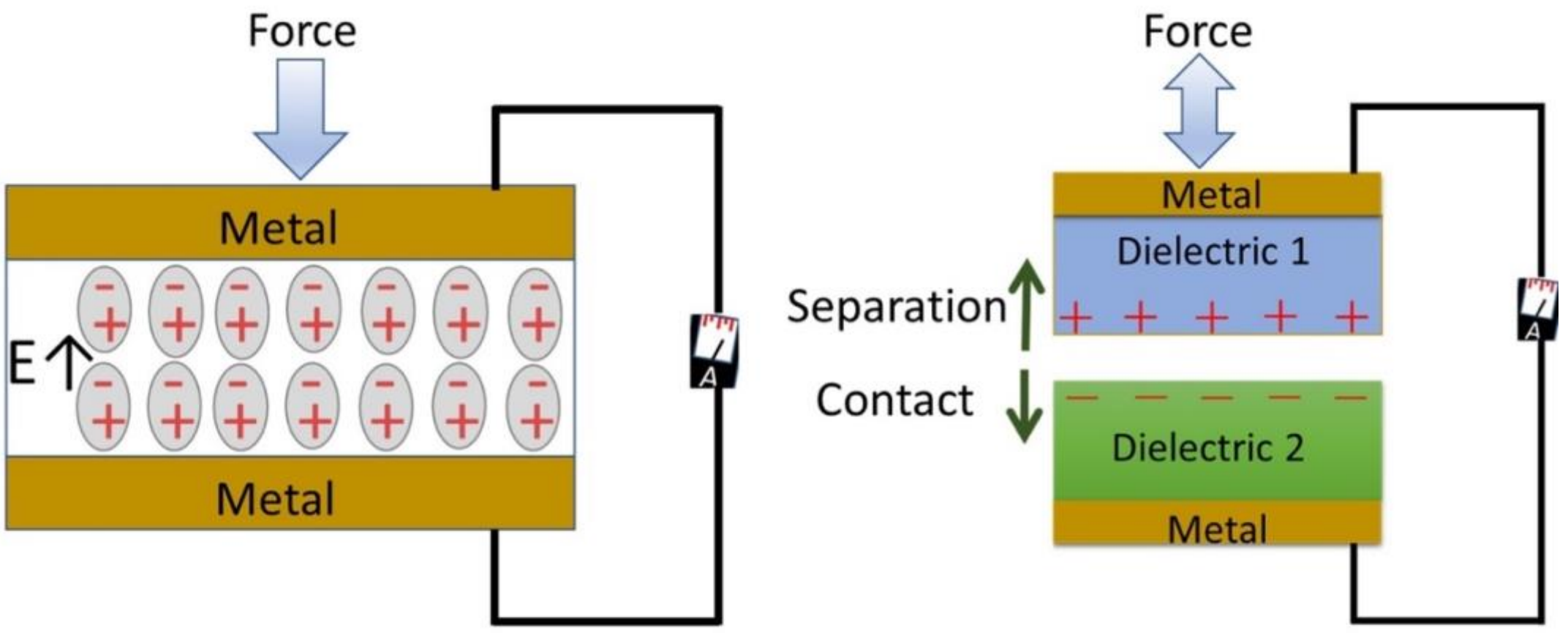

a) Piezoelectric b) Triboelectric

**Figure 4**

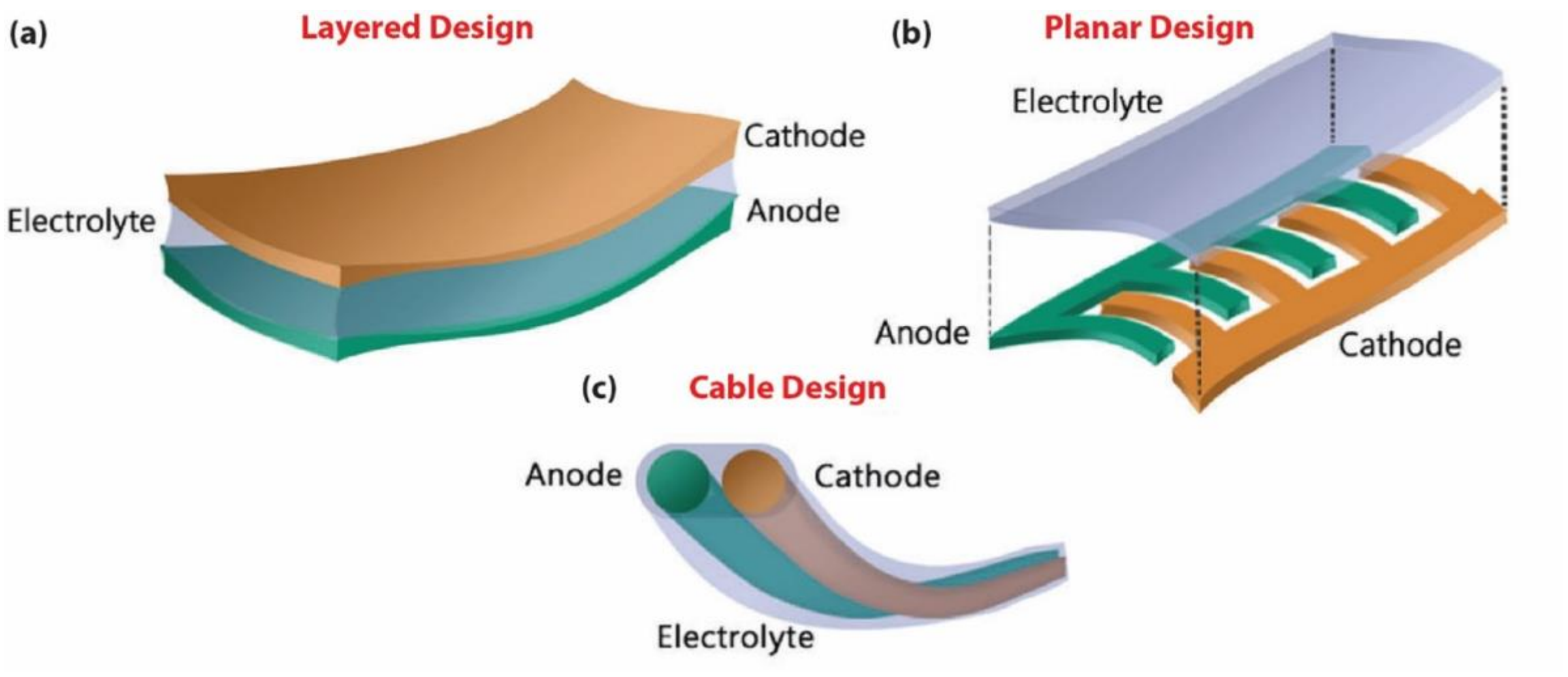

**Figure 5**

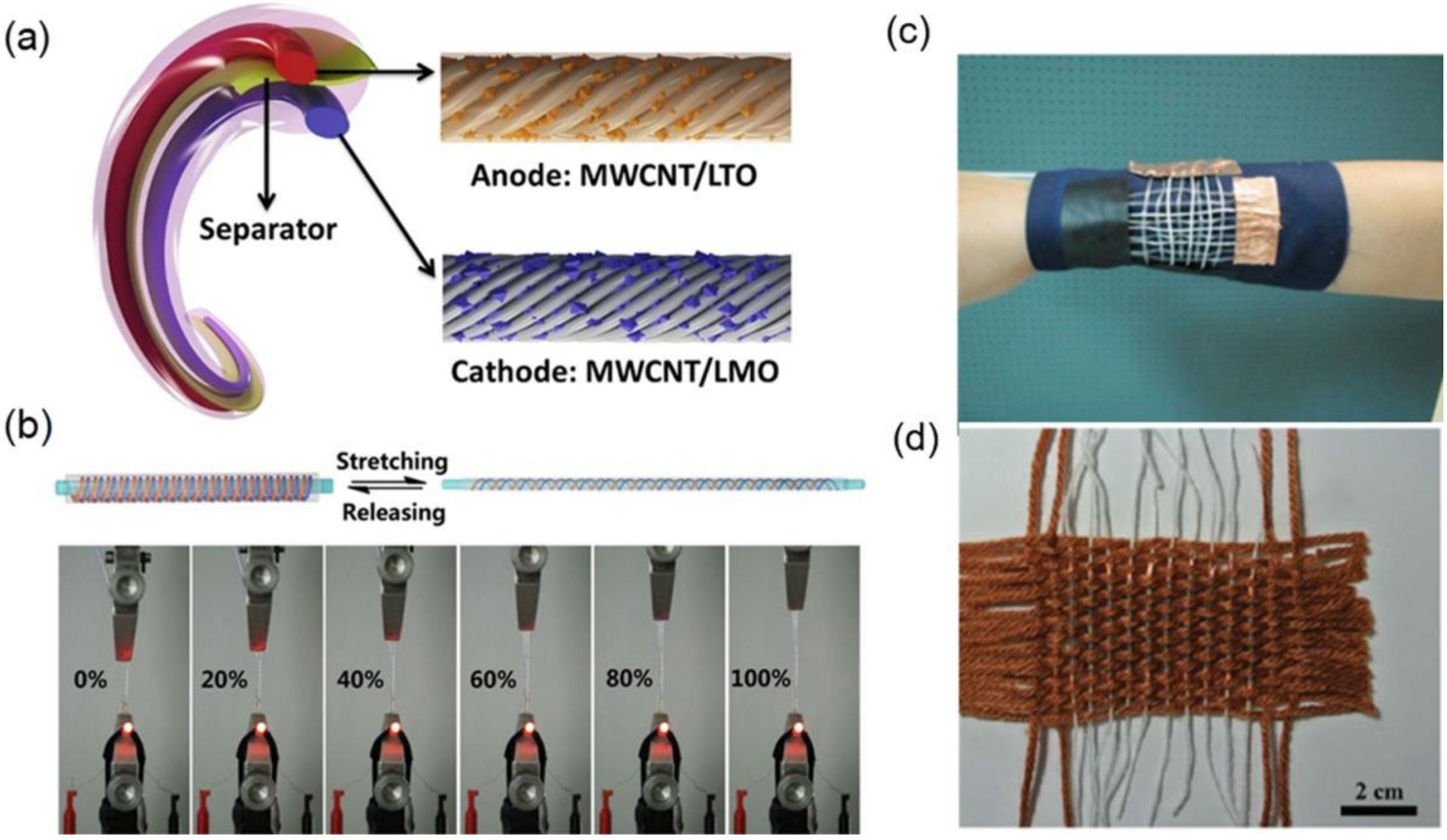

**Figure 6**

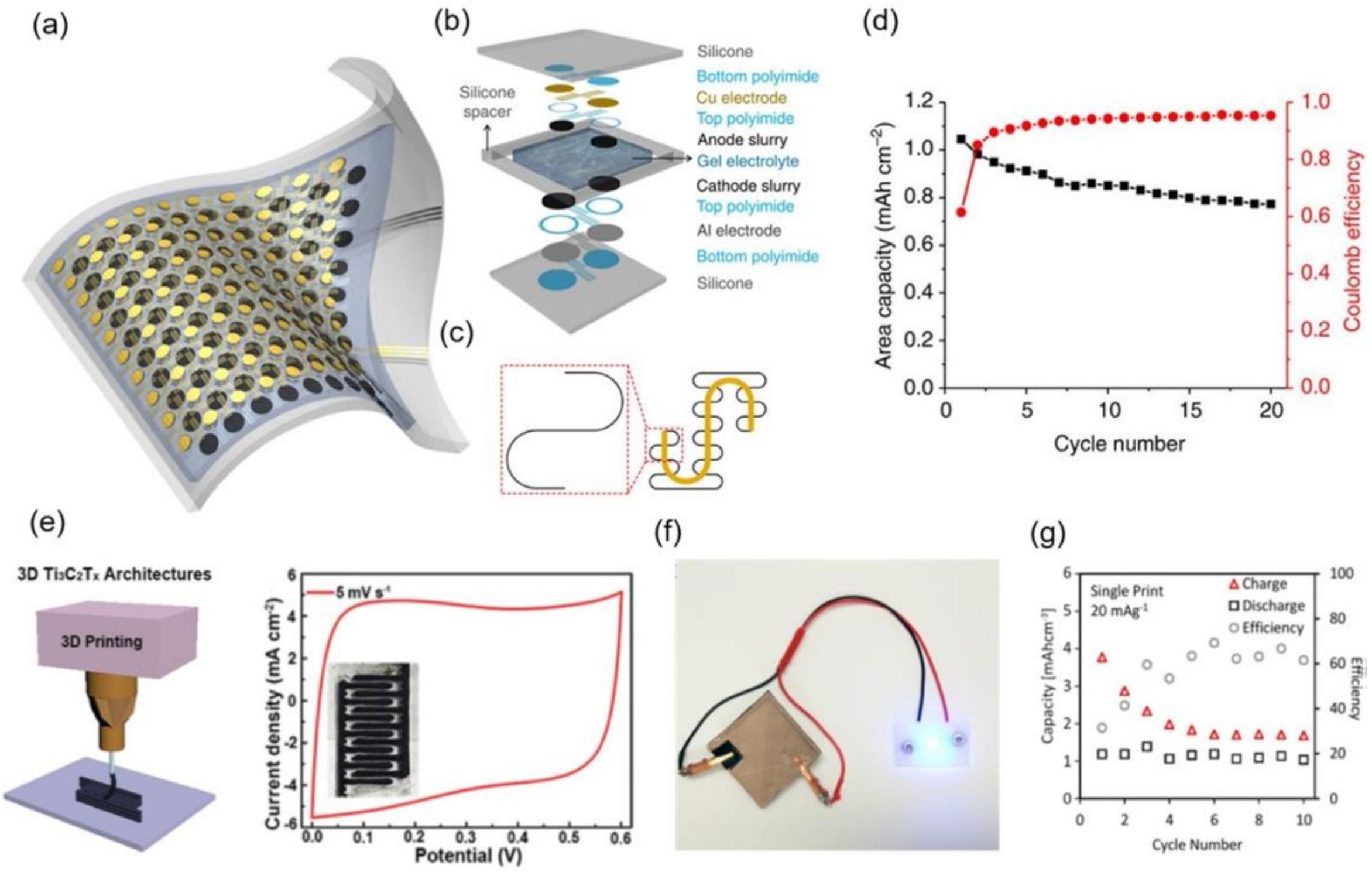

**Figure 7**

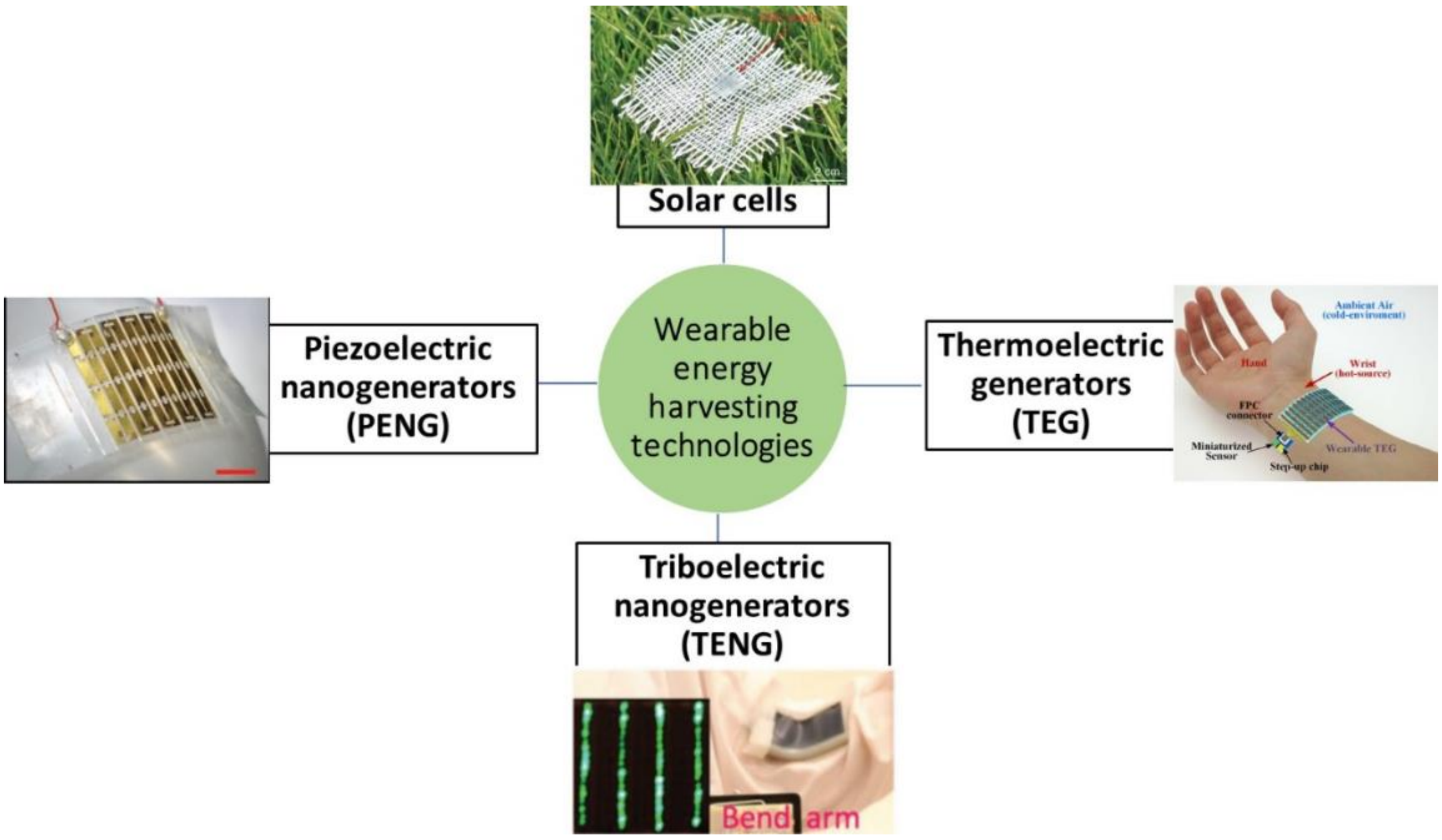

**Table 1**

| Sensing technology | Active sensing material | Device Location | Sensitivity | Flexible/ Stretchable | Application |
|---|---|---|---|---|---|
| Piezoresistive[20] | carbon black-decorated fabric | Chest and wrist | 0.585 $kPa^{-1}$ | Flexible | Pulse and ECG sensor |
| Piezoresistive[77] | SWCNT | Wrist | 1.80 $kPa^{-1}$ | Flexible | Monitoring human physiological signals such as wrist pulse and muscle movement |
| Piezoresistive[61] | Ag NWs / rubber | Finger and wrist | 4.29 $N^{-1}$ | Stretchable | Monitoring human pulse wave and body posture and movement |
| Piezoresistive[78] | Au NWs / rubber | Fingers, arms and neck | GF = 9.9 | Stretchable | Monitor human body movements |
| Piezoresistive[79] | Graphene foam / PDMS | Wrist and fingers | GF = 15 - 29 | Stretchable | Monitoring elbow and fingers bending and pulse of radial artery |
| Piezoresistive[80] | Reduced Graphene oxide (rGO) and PVDF | Wrist | None available | Flexible | Monitoring of artery pulse pressure |
| Piezoresitive[37] | Au NWs | Wrist | >1.14 $kPa^{-1}$ | Flexible | Monitoring of blood pulses |
| Piezoresistive[81] | CNT/rGO/ZnO NW | Elbow | GF =7.64 | Stretchable | Wireless monitoring of elbow bending |
| Piezoresistive[82] | Graphene/rubber | Finger, wrist and neck | GF =35 | Stretchable | Monitoring of finger bending, forearm movement, phonation, breathing, and pulse |
| Piezoresistive[83] | SWCNT/PU-PEDOT:PSS | Face | GF = 8.7–837.1 for 1.5–3.5% | Stretchable | Monitoring of facial muscle movement |
| Piezoresistive[84] | SWCNT/self-healing hydroge | Different parts of the human body | GF = 0.24 (100%), 1.51 (1000%) | Stretchable | Monitoring of finger/knee/neck/elbow bending |
| Piezoresistive[57] | CNT/PDMS | Neck, knee and fingers | GF = 0.82 (0–40%), 0.06 (60–200%) | Stretchable | Data glove, monitoring of knee motions, breathing, and phonation |
| Piezoresistive[85] | Interlocked CNT/PDMS microdome array | Finger, medical breathing mask, and neck | 15.1 $kPa^{-1}$ (<0.5 kPa) | Flexible | Monitoring of finger bending, breathing, phonation, |
| Piezoresistive[86] | Carbonized silk nanofibe | Neck, wrist and chest | 34.47 $kPa^{-1}$ (0.8–400 Pa), 1.16 $kPa^{-1}$ (400–5000 Pa) | Flexible | Monitoring of wrist pulse, chest respiration, and phonation |
| Capacitive[73] | AgNW/Ecoflex/ AgNW | Finger and knee | GF = 0.7 Pressure sensitivity = 1.62 $MPa^{-1}$ | Stretchable | Monitoring of finger bending and knee motion |
| Capacitive[76] | CNT/Dragon Skin/CNT | Fingers | GF = 1 | Stretchable | Date glove, monitoring of balloon inflation and chest movement |
| Capacitive[87] | CNT/porous PDMS/air gap/CNT | - | 0.7 $kPa^{-1}$ (0–1 kPa), 0.14 $kPa^{-1}$ (1–5 kPa), 0.005 $kPa^{-1}$ (5–20 kPa) | Stretchable | Electronic skin |
| Capacitive[66] | AgNW /AgNW–PU/ AgNW | Finger, arms and knee | 5.54 $kPa^{-1}$ (0–30 Pa), 0.88 $kPa^{-1}$ (30–70 Pa) | Flexible | Monitoring of knee/finger bending, forearm muscular movement and air blow |

**Table 2**

| Sensing technology | Active sensing material | Device Location | Sensitivity | Flexible/ Stretchable | Application |
|---|---|---|---|---|---|
| PENG[97] | PVDF film | Pig heart and human-knee | None available | Stretchable | Health monitoring |
| PENG[96] | PZT fibers | Human heart, lungs and diaphragm | None available | Flexible | Mechanical energy harvesting |
| PENG[110] | PVDF-TrFE NWs | Finger and wrist | 0.45 V $N^{-1}$ | Flexible | Finger movements and health monitoring |
| PENG[93] | PVDF-TrFE nanofiber | Finger and arms | 0.4–0.8 V $kPa^{-1}$ | Flexible | Accelerometer, orientation sensor |
| PENG[111] | PVDF-TrFE nanofiber | Wrist | 0.3 V $kPa^{-1}$ | Flexible | Monitoring of radial and carotid pulses |
| PENG[95] | ZnO NW-parylene hybrid | Finger | 0.09 V $kPa^{-1}$ | Stretchable | Finger motion monitoring |
| TENG[104] | Kapton and Cu | carotid artery | None available | Flexible | Monitoring arterial pulse wave |
| TENG[112] | PTFE and PET | carotid artery | 0.05 V $Pa^{-1}$ | Flexible | Monitoring arterial pulse wave |
| TENG[113] | FEP and Au | Ears | 0.11 mV $dB^{-1}$ | None | Hearing aids |
| TENG[114] | PET and PDMS | Finger | 0.0323 $N^{-1}$ | Flexible | Finger motion monitoring |
| TENG[115] | FEP and Acrylic | Finger | None available | Flexible | Human-robotic interface and finger motion monitoring |
| TENG[116] | Rubber and Al | Stomach | None available | Stretchable | Respiration monitoring |
| TENG[117] | PET and rubber | Insole | 0.8 V $N^{-1}$ | None | Gait monitoring |
| TENG[118] | Acrylic and PTFE | Medical mask | None available | None | Respiration monitoring |

**Table 3**

| Analytes in sweat | Electrochemical sensing method | Receptor element | Sensing sensitivity and characteristic | Flexible/ Stretchable | Ref. |
|---|---|---|---|---|---|
| $Na^+$ | Potentiometric | $Na^+$ sensitive PVC and Ag/AgCl ink reference | 110 ± 5 mV/pNa for $Na^+$ [10 – 100 mM] | Flexible | 129 |
| | Potentiometric-based ISFET | | 55 mV/pNa good selectivity toward $K^+$ potassium ion no pH interferences in the range [2–9] | | 129 |
| $Na^+$ | Potentiometric | $Na^+$ Ionophore and Ag/AgCl on CNT | 45.8 mV $dec^{-1}$ for $Na^+$ [10–160 mM] | Flexible | 147 |
| $Cl^-$ | Conductivity | - | Single use patch Working range: [5 – 120 mM equivalent NaCl] Power: Max 2 mW | Flexible | 132 |
| $K^+$ | Potentiometric | $K^+$ Ionophore and Ag/AgCl on CNT | 35.9 mV $dec^{-1}$ for $K^+$ [2–32 mM] | Flexible | 147 |
| $Ca^{2+}$ | Potentiometric | $Ca^{2+}$ Ionophore and Ag/AgCl on CNT | 52.3 mV $dec^{-1}$ for $Ca^{2+}$ [0.5–2.53 mM] | Flexible | 147 |
| pH | Potentiometric | PANI and Ag/AgCl screen printed on paper | Tattoo like sensor pH [3–7] Sensitivity: 51 mV/pH Sensitivity by stretching: 58mV/pH | Flexible | 127 |
| pH | Potentiometric | PANI on CNT and Ag/AgCl on CNT | pH [4–7] Sensitivity: 40 mV/pH | Flexible | 147 |
| Glucose | Amperometric | Glucose oxidase mixed with agarose | Glucose range: [15–200 µM] | Flexible | 148 |
| Lactate | Amperometric | Lactate oxidase | Sensitivity of 0.16 nA $\mu M^{-1}$ | Flexible | 137 |
| Cortisol | Impedemitric | Cortisol anti-body on Gold electrodes on Polyamide. | Cortisol in sweat: [10–200 ng/mL] pH [4–8] Temp. [25–40°C] response time: 13 ms | None | 149 |
| Cortisol | Impedemitric | Cortisol-antibody functionalized $MoS_2$ nanosheets deposited into porous polyamide membrane fixed between two Pt electrodes | Cortisol in sweat: [1–500 ng/mL] Freq. 100 KHz prototype | None | 145 |
| Cortisol | Amperometric | Cortisol anti-body is covalently bonded to $F_2O_3$ on carbon fiber | Cortisol in sweat: [$10^7$–100 ng/mL] | Flexible | 133 |

**Table 4**

| Design | Anode/CC | Cathode/CC | Electrolyte | Performance | Ref. |
|---|---|---|---|---|---|
| Planar | C/Cu | NCA/Al | $LiPF_6$/PVDF-HFP | 962 W h $kg^{-1}$ and 0.055 kW $kg^{-1}$ (30 cycles) at bending angle of $120^0$ | 189 |
| Planar | Zn | G-$MnO_x$-CC/$O_2$ | KOH/PAA | 44.1 mW $cm^{-2}$ (145 cycles) | 190 |
| Layered | LTO/Cu | LCO/Al | $LiClO_4$/PEO | 2.53 mW h $cm^{-2}$ and 4.8 mW (20 cycles) at 300% strain | 188 |
| Layered | Graphite/Cu | LCO/Al | $LiPF_6$ | 160 W h $kg^{-1}$ and 2.75 kW $kg^{-1}$ (100 cycles) at bending angle of $180^0$ | 191 |
| Layered | Zn/Cu | CNT/$LaNiO_3$/$O_2$ | KOH/PVA | 581 W h $kg^{-1}$ and 0.06 kW $kg^{-1}$ (120 cycle) at bending angle of $180^0$ | 156 |
| Cable | LTO/CNT | LMO/CNT | LiTFSI-PEO | 27 W h $kg^{-1}$ and 0.88 kW $kg^{-1}$ (200 cycles) at 100% strain | 192 |
| Cable | Zn | $Co_3O_4$/r-GO/$O_2$ | KOH/PVA | 36.1 mW h $cm^{-3}$ and 7.08 mW $cm^{-3}$ (75 cycles) at bending angle of $180^0$ | 193 |
| Cable | Na/Cu | Prussian Blue-rGO/Ni | $NaPF_6$ | 176 W h $kg^{-1}$ and 0.14 kW $kg^{-1}$ (500 cycles) at bending angle of $90^0$ | 194 |

**Table 5**

| Design | Fabrication | Active material/s | Electrolyte | Performance | Ref. |
|---|---|---|---|---|---|
| Planar | Screen printing | $MnO_2$-carbon/Paper | $H_3PO_4$/PVA | 7.04 $mFcm^{-2}$ | 163 |
| Planar | Screen printing | CNT/PC | TMOS:FA:EMI TFSI | 430 $\mu Fcm^{-2}$ | 197 |
| Planar | Inkjet printing | Graphene/Si | $H_3PO_4$/PVA | 0.1 $mFcm^{-2}$ | 198 |
| Layered | Screen printing | $MnO_2$-graphene/PET | 2 M Ca $(NO_3)_2$ | 175 F $g^{-1}$ | 161 |
| Layered | R2R printing | Carbon slurry | $H_3PO_4$/PVA | 45 $mFcm^{-2}$ | 164 |
| Cable | Parallel | C/Carbon fiber | $H_2SO_4$/PVA | 4 mF $cm^{-2}$ | 171 |
| Cable | Core-shell | CNT/CNT sheet | $H_3PO_4$/PVA | 0.029 mF $cm^{-1}$ | 170 |
| Cable | Intertwined | $MnO_2$/CNT fiber | KOH/PVA | 25.4 F $cm^{-3}$ | 169 |